\documentclass[amsmath,amssymb,showkeys,twocolumn,pre]{revtex4}

\usepackage{graphicx}
\usepackage{subfigure}
\begin{document}

\title{Distance Measures for Dynamic Citation Networks}

\author{Michael J. Bommarito II (mjbommar@umich.edu)}
	\affiliation{Department of Political Science, University of Michigan, Ann Arbor}
	\affiliation{Department of Mathematics, University of Michigan, Ann Arbor}
	\affiliation{Center for the Study of Complex Systems, University of Michigan, Ann Arbor}

\author{Daniel Martin Katz (dmartink@umich.edu)}
	\affiliation{University of Michigan Law School}
	\affiliation{Department of Political Science, University of Michigan, Ann Arbor}	
	\affiliation{Center for the Study of Complex Systems, University of Michigan, Ann Arbor}

\author{Jon Zelner (jzelner@umich.edu)}
	\affiliation{Department of Sociology, University of Michigan, Ann Arbor}
	\affiliation{Gerald R. Ford School of Public Policy, University of Michigan, Ann Arbor}
	\affiliation{Center for the Study of Complex Systems, University of Michigan, Ann Arbor}

\author{James H. Fowler (jhfowler@ucsd.edu)}
	\affiliation{Department of Political Science, University of California, San Diego}
	\affiliation{Center for Wireless and Population Health Systems, University of California, San Diego}

\date{\today}

\begin{abstract}
\textbf{Abstract:} Acyclic digraphs arise in many natural and artificial processes.  Among the broader set, dynamic citation networks represent a substantively important form of acyclic digraphs.  For example, the study of such networks includes the spread of ideas through academic citations, the spread of innovation through patent citations, and the development of precedent in common law systems.  The specific dynamics that produce such acyclic digraphs not only differentiate them from other classes of graphs, but also provide guidance for the development of meaningful distance measures.  In this article, we develop and apply our sink distance measure together with the single-linkage hierarchical clustering algorithm to both a two-dimensional directed preferential attachment model as well as empirical data drawn from the first quarter century of decisions of the United States Supreme Court.  Despite applying the simplest combination of distance measures and clustering algorithms, analysis reveals that more accurate and more interpretable clusterings are produced by this scheme.
\end{abstract}

\maketitle

\section{Introduction \& Motivation}
While a variety of algorithms exist for the analysis of undirected or cyclic graphs, e.g., social networks, comparatively little work has been done on acyclic digraphs. Previous literature has focused particularly on the development of canonical random graph models or the application of algorithms for general graphs to this special class (\cite{Karrer2009}, \cite{Bollobas2003}, \cite{Bommarito2009}). While these initial papers have made important contributions, additional investigation of dynamic acyclic digraphs (DADGs) still remains. 
\par
Dynamic acyclic digraphs arise naturally in the context of document citation networks.  In these networks, nodes represent documents and arcs represent the citations from one document to another.  Much of the previous literature on citation networks, however, disregards the direction of these arcs (for a notable exception see Leicht, et al., \cite{Leicht2007}).  This choice results in undirected graphs with many cycles, and thus allows the application of a wide variety of well-developed algorithms.  On the other hand, disregarding direction does discard information about time and the flow of dependency.
\par
Recent work demonstrates that applying methods for undirected cyclic graphs to citation networks may create difficulties (\cite{Bommarito2009}).  While it is important to identify deficiencies in existing methods, it is more helpful to develop alternative approaches designed to properly address these shortfalls. With respect to the context in question, we seek to develop domain-specific methods and measures for citation networks that take the acyclic digraph nature of these networks into account.  In this article, we present a novel sink distance measure that provides better computational efficiency and qualitative accuracy than traditional measures.  

\section{Properties of Dynamic Citation Networks}
\subsection{Topological Ordering}
Since citation networks are dynamic acyclic digraphs, they feature a number of important properties that distinguish them from other networks.  The most fundamental property of acyclic digraphs is that there exists at least one topological ordering of the nodes (\cite{BangJensen2000}). Such a topological ordering can also be used to index the dynamic network $G$ itself, where $G$ is a nested set of increasing graphs $\{G_1, G_2, \dots, G_{|V|}\}$.  Each $G_n$ is a copy of $G_{n-1}$ with the addition of the $n^{th}$ document of the topological ordering and its corresponding arcs. From this growth dynamic, it is clear that the most natural topological ordering is actually the chronological ordering of the documents. 
\par
This ordering implies the existence of another distinguishing property for this class of graphs.  Unlike many other growing networks, the set of arcs with non-zero probability at each time step can be explicitly constrained.  From a generative framework, this representation acknowledges that arcs cannot assert relationships with unobserved nodes at later times \footnote{Draft circulation and pre-print repositories such as arXiv or SSRN allow for the existence of multiple versions of a given document.  Given differential delays between drafts and subsequent publication it is possible, at least in theory, for cycles to exist. It is important for a researcher to consider how best to represent documents that are not consistent with the filtration.}.  Formally, such a process evolves on a filtration and can sample from the set of possible arcs at time $t$ given by $\Omega_t^A = \{(x,y) : x \notin V(G_{t-1}), y \in V(G_{t-1})\}$, where $V(G_t)$ is the set of vertices in the graph $G_t$ and $t$ is the index corresponding to the topological ordering\footnote{While we define a degree-agnostic constraint, the study of many citations practices indicate a self-organized convergence upon a highly skewed degree distribution.}.  From a statistical framework, in which only the resulting graph is observed, the previous statement can be written $T(x) \leq  T(y) \Leftrightarrow \mathbb{P}((x,y) \in A(G)) = 0$, where $T(x)$ is the time that node $x$ was introduced into the graph $G$.  This asserts that certain events should not even be considered as possible in statistical models.

\par
\subsection{Sinks and Dimensionality}
A fact that follows immediately from the existence of a topological ordering is that there is at least one document that makes no citations and at least one document that has never been cited. Documents that contain no citations correspond to nodes with out-degree zero and are called ``sinks."  The first node in the topological ordering is always a sink. Sinks represent documents with no dependencies observed. Thus, with respect to the data provided, they mark the introduction of at least one original or novel idea(\footnote{An allied approach could be applied to capture the diffusion of ideas across an author to author projection of the network.}). Nodes that are not sinks then rely on one or more of the ideas provided in one or more sinks. 
\par
Though the above conception of citation networks is simple and reasonable, it contradicts patterns often observed in empirical citation data. Namely, many documents contribute novel ideas, but very few feature zero outbound citations. In order to confront this complication and refine our initial conception, it is important to remember documents and citations exist in a high-dimensional space. Documents may contribute novel ideas in one dimension but draw support or comparison from other dimensions - we call these documents ``weak'' sinks, as opposed to ``strong'' sinks which make no citations in any dimension. 
\par
For a simple but concrete depiction of this problem, consider \textit{Figure 1} below, a hypothetical subgraph containing nodes $a$, $b$, and $c$ respectfully. Node $a$ is a ``strong'' sink as it features no outbound citations.  Node $b$ is a ``weak'' sink as it cites $a$ on the red dimension but generates no citations with respect to its blue dimension. Node $c$ is not a sink, as it relies on $b$ and does not contain the red dimension. 

Lacking appropriately granular data, it is often difficult for researchers to separate the dimensions contained within the observable outputs of a given system. However, the above example highlights the specific usefulness of dimensional data. It is important to note that dimensional data is only necessary to identify ``weak'' sinks but not ``strong'' sinks. For example, if dimensional data were removed from \textit{Figure 1} below, thereby removing the coloring of the subgraph, only node $a$ would be identified as a sink.\footnote{The added returns to differentiating strong and weak sinks will likely vary between substantive problems.  However, all else equal, if reasonably well specified dimensional data for arcs and/or nodes were available, it would probably enhance the quality of the subsequent analysis.}

\begin{figure}[htp]
	\includegraphics[width=3.5cm]{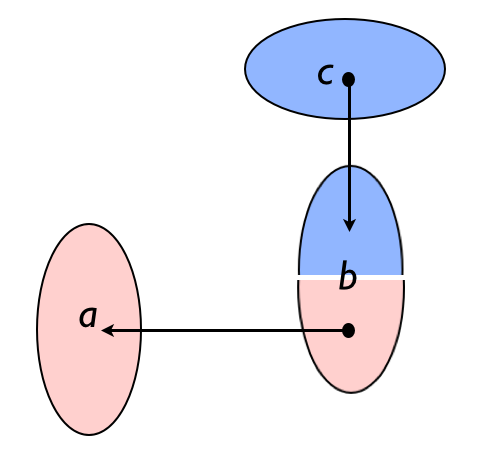}
	\caption{Dimensional Data and ``Weak'' vs ``Strong'' sinks}
\end{figure}

\par
In the context of acyclic digraphs, consider a citation network comprised of linkages between academic articles. While citations to a given article often converge upon a particular dimension or aspect of the work, a given article could be cited on the basis of any of its $n$ dimensions.  Building from the example offered in $Figure$ $1$, assume node $b$ is an article containing both a novel method and an interesting substantive result. \footnote{While a given article may contain multiple methods or multiple substantive results, for simplicity, assume an article containing a single methodological contribution and a single substantive contribution.}  If the \textit{blue dimension} represents the article's substantive topic while the \textit{red dimension} represents the article's methodological contribution, then the basis upon node $c$ cites node $b$ and $b$ cites node $a$ could only be definitively revealed with dimensional data.        

While this example is trivial, it reveals a broader property of acyclic graphs. While an author can select any citation from the existing citation set, that author has no specific control over the basis upon which his or her work is subsequently cited. One positive feature of the sink method we offered herein is that it generally preserves the choices made by the author at the time of authorship.  

\section{Distance Measures}
Distance measures between nodes are the predicate to a wide variety of algorithms in machine learning.  As noted earlier, we believe that distance measures employed should incorporate the properties of dynamic citation networks that differentiate them from other classes of graphs.  We consider the distance between nodes in the ``citation'' space, where all documents must orient themselves relative to one or more sinks of information.

An appropriate distance measure should decrease as two nodes share more information.  The simplest such measure should consider the number of shared sinks between two nodes.  Given a node $i$ and its set of ancestors $A_i$, the sinks of $i$ are given by the set $S_i = \{x : \delta^+(x) = 0, x \in A_i\} = S \cap A_i$.  Here, $\delta^+(x)$ is the standard notation for the out-degree of node $x$ and $S$ is the set of all sinks of the graph $G$.

Using this notation, we can represent the distance between nodes $i$ and $j$ as the proportion of sinks they do not share:
\begin{align}
	D_{i,j} =& 1 - \frac{|S_i \cap S_j|}{|S_i \cup S_j|}
\end{align}
where $|x|$ is the cardinality of set $x$.  Though this distance measure is linearly decreasing in the proportion shared, one can formulate a distance measure from any appropriately decreasing function.  The remainder of our distance measures will feature this linear form, but the reader should keep in mind that this is only exemplary.

Furthermore, this measure can be calculated quickly for all pairs of nodes, as its implementation involves little more than graph traversal and set operations.

In the above distance measure, we weight all sinks equally.  An alternative measure might weight the importance of each sinks by the number of unique ancestors shared between nodes $i$ and $j$ that are descended from a sink $s$ of interest.  This set of $s$-ancestors of node $i$ is given by $A_{i,s} =\{x : s \in S_x, x \in A_i\} = A_i \cap D_s$, where $D_s$ is the set of descendents of $s$.  This can be interpreted as the ancestors of $i$ who carry the information from sinks $s$.  If even more detail is desired, we might modify the above measure to also incoporate the set of paths from nodes $i$ and $j$ to the sinks of interest.  We let $P_{s,i}$ be the set of path tuples $(x_1,\dots,x_n)$ from $s$ to $i$. The resulting general equation takes the form
\begin{align}
	D_{i,j} =& 1 - \frac{\sum_{s \in S_i \cap S_j} f(A_{i,s}, P_{i,s}, A_{j,s}, P_{j,s})}{\sum_{s \in S_i \cup S_j} f(A_{i,s}, P_{i,s}, A_{j,s}, P_{j,s})}
\end{align}

Straightforward choices of $f$ involve the cardinality of these sets, but some care must again be taken if one desires a distance metric that obeys all axioms.  If needed, these functions may take on much more complexity.  For instance, the importance of a sink might decay as its shortest path length increases.  Such fine-grained choices however require substantive justification drawn from the given problem at hand.  One should also note that path-based algorithms are likely to be more complex than either of the first two measures.

The above distance measures all bear some similarity to the Jaccard similarity measure, as they involve intersections in the numerator and unions in the denominator (\cite{Jaccard1901}).  However, the standard Jaccard similarity index only takes into account the neighbors of each node and ignores nodes of any further distance.  Thus, the above distance measures may better capture and weight ``shared ancestry'' than the Jaccard similarity measure.

\section{Applications}
Once a distance measure has been offered, a number of interesting research questions become relevant.  One question of particular interest is whether a given graph exhibits detectable clustering or community structure.  A significant amount of recent scholarship has been devoted to the production of community detection algorithms for general graphs (\cite{Porter2009}). However, as noted earlier, in the context of acyclic citation networks, there are a number of issues that may impact the both accuracy and time evolving stability of substantive results returned by traditional community detection methods.  

One important issue is dimension frequency - that is, some topics may occur much more frequently than others in the overall network.  For example, suppose that a node $z$ primarily concerns dimension $d_1$, but also touches upon dimension $d_2$. If subsequent documents more often confront dimension $d_2$ than $d_1$, it is possible that $z$ could receive more $d_2$-related citations than $d_1$ citations.  As a result, traditional community detection methods are more likely to cluster document $z$ with $d_2$-related documents than with $d_1$ documents.  Though this example illustrates the evolutionary nature of documents within such citation networks, it is clear that traditional community detection algorithms may produce clusterings that differ from a given researcher's goals.

Instead, one might seek to cluster documents in a manner consistent with the citation choices of the author at the time the document was written.  In this case, sink-based distance measures as presented above might be a good choice for clustering.  Take the above node $z$ as an example.  Suppose a node $z$ has three sinks linked to dimension $d_1$, but only one sink dealing with dimension $d_2$.  Even if many more $d_2$ documents cite $z$, they can only share one of four sinks at most.  By contrast, $d_1$ documents can share up to three sinks with $z$.  Thus, regardless of the number of citations from $d_1$ and $d_2$ documents, $z$ can still be  closer to $d_1$ documents.  Though many gradations of this example exist, when confronted with unnormalized and high-dimensional citation information, sink-based distance measures are likely to be more robust to this issue than traditional community detection methods.

To test whether meaningful clustering can be derived from these sink distance measures, we apply \textit{equation (1)} to two networks below, a theoretical model generated by two-dimensional directed preferential attachment and the other from substantive data offered in the citations of the early United States Supreme Court.

\subsection{Comparison on a Random Model}
In \textit{Section II.B} above, we argue that a number of issues can cause problems with existing community and clustering algorithms.  To test this claim, we have generated realizations from a citation model based on two-dimensional directed asymmetric preferential attachment (\cite{Barabasi1999}).

The model has two types of nodes - red and blue.  At each model step, a new node is introduced into the network.  With probability $l_r$, a node will be red, and thus the complement $l_b$ is the probability of the node being blue.  To determine how many citations this node will make, we sample a uniform random integer between $1$ and $m$.  These citation edges are assigned according to the directed preferential attachment model, where red nodes have probability $p_{rr}$ of citing red nodes and probability $p_{rb}$ of citing blue nodes.  Likewise, blue nodes have probability $p_{br}$ of citing red nodes and probability $p_{bb}$ of citing blue nodes.  For initial conditions, there are a $n_r$ initial red nodes and $n_b$ initial blue nodes.  

In order to demonstrate the problems described above, we choose the parameters of the model to emphasize our example from Section \textit{II.B}.  The node type rates are given by $l_r = \frac{1}{4}, l_b = \frac{3}{4}$, the maximum number of edges per node is given by $m = 3$, the preferential probabilities are given by $p_{rr} = 1, p_{br} = \frac{1}{4}, p_{bb} = \frac{3}{4}$, and the initial number of nodes of each type are given by $n_r = 2, n_b = 1$.  This models a system with two dimensions where each node may only have one dimension, and one dimension occurs much more frequently than another.  Furthermore, though one dimension is perfectly homophilic, the other attaches to both.  Figure 2 below shows an example realization of this model, where the large squares denote sinks.

\begin{figure}[htp]
	\includegraphics[width=8cm]{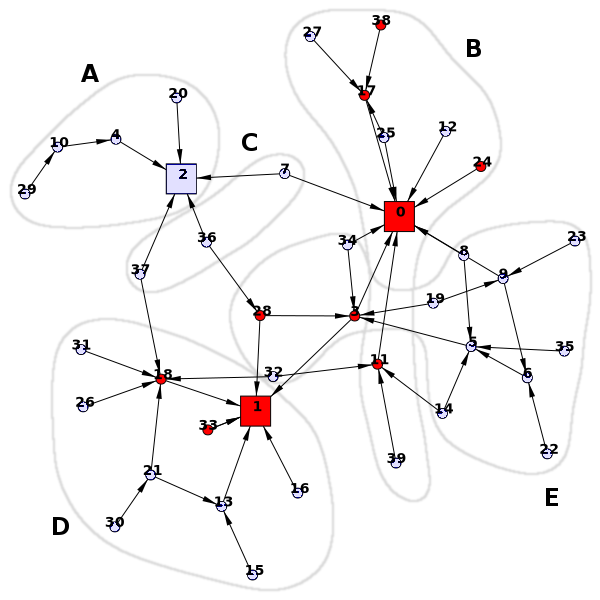}
	\caption{Realization of Random Model}
\end{figure}

To justify our method, we compare our sink-based approach with directed edge-betweenness (\cite{Newman2004}).  First, we apply \textit{equation (1)} to calculate a full distance matrix for all pairs of nodes.  Using this matrix,  we then apply a single-linkage hierarchical clustering algorithm to these distances (\cite{Murtagh1983}).  The resulting dendrogram and its implied clustering are shown in Figure 3(a) and Figure 2, respectively.  Next, we apply the directed edge-betweenness algorithm to produce the merge dendrogram in Figure 3(b).  Both Figures 2 and 3 are generated from the same underlying network visualized in Figure 2.  The numbering on both figures corresponds to the nodes, and the letters A through E correspond to the clusters detected by the sink method in Figure 3(a).

\begin{figure}[htp]
	\subfigure[\ Sink]{\includegraphics[width=4cm]{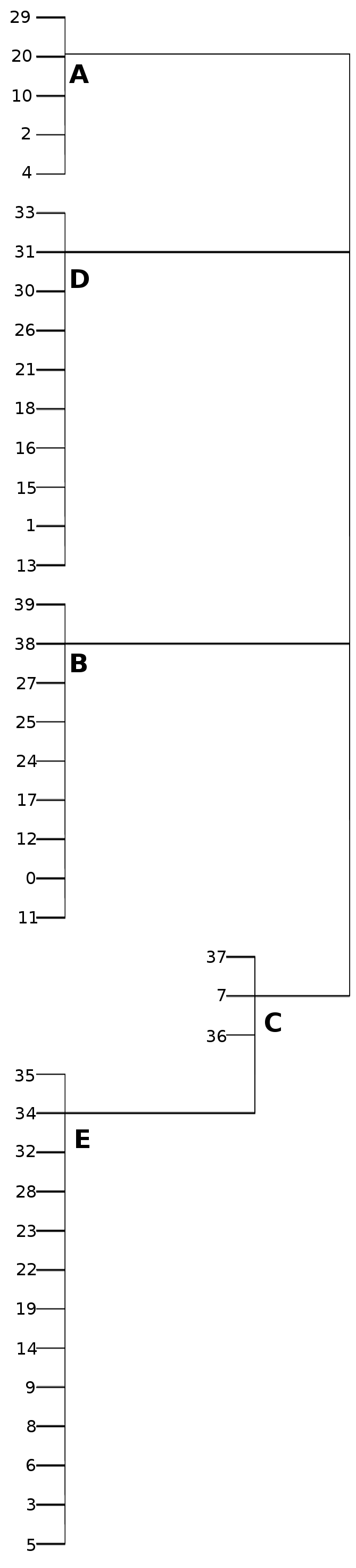}}
	\subfigure[\ Betweenness]{\includegraphics[width=4cm]{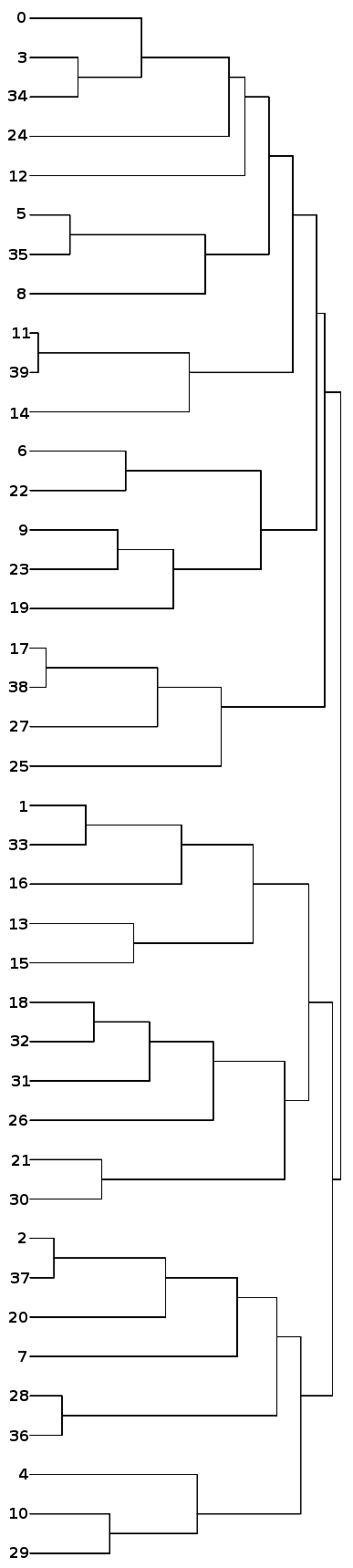}}
	\caption{Clustering Dendrograms}
\end{figure}

The differences in Figure 3 are striking, though not entirely unexpected.  The sink method in Figure 3(a) identifies five ``communities'' of nodes at identical branch location.  From top to bottom, these branches correspond to (1) nodes that trace back only to node 2, (2) nodes that trace back only to node 1, (3) nodes that trace back only to node 0, (4) nodes that trace back to all three sinks 0, 1, and 2, and (5) nodes that trace back to both nodes 0 and 1.  

Since the edge-betweenness algorithm produces binary branching dendrograms like most agglomerative and divisive algorithms, Figure 3(b) exhibits a great deal more complexity than Figure(a). This complexity is sometimes warranted; however, it is often the product of ties in the agglomerative or divisive decision criteria.  Since the sink method relies only on hierarchical clustering, it places nodes with equal distance at an equal branch position.  In this case, the sink method identifies clusters that are closely related to the underlying network formation process.

\subsection{Results for the United States Supreme Court Citations}
To generate applicability beyond the context of a theoretical model, we applied our approach to the case-to-case citation network contains the first quarter century of decisions of the United States Supreme Court. The structure of this network of citations is of interest to a wide variety of scholars including not only law professors and social scientists, but also members of the physical science community (\cite{Smith2007}, \cite{Fowler2007}, \cite{Leicht2007}, \cite{Post2000}).  While it is possible to perform community detection analysis over the total body of Supreme Court decisions, we selected a reduced window of decisions in order to qualitatively examine the results of our algorithm (\footnote{With respect to the application of clustering or community detection methods, a number of recent articles have called for more extensive qualitative validation of detected outputs (\cite{Porter2009}) Accepting this call, we substantively vetted the outputs generated by our method.} 

The Court's early citation practices indicate an absence of references to its own prior decisions.  While the court did invoke well-established legal concepts, those concepts were often originally developed in alternative domains or jurisdictions \footnote{The Supreme Court's early jurisprudence references the decisions of England and France as well as several state courts.}.  At some level, the lack of self-reference and corresponding reliance upon external sources is not terribly surprising.  Namely, there often did not exist a set of established Supreme Court precedents for the class of disputes which reached the high court.  Thus, it was necessary for the jurisprudence of the United States Supreme Court, seen through the prism of its case-to-case citation network, to transition through a loading phase.  During this loading phase, the largest weakly connected component of the graph generally lacked any meaningful clustering.  However, this sparsely connected graph would soon give way, and by the early 1820's, the largest connected component displayed detectable structure.  

Despite applying the most naive assumptions and least complicated clustering algorithm, our qualitative analysis reveals that accurate clusterings are produced by this scheme. By applying our sink clustering method, we obtain a dendrogram of the network's largest weakly connected component shown in \textit{Figure 2}. The coloring in both \textit{Figure 2} and \textit{Figure 3} corresponds to two large clusters in the network. Edges are colored blue or red if the head or tail are of the respective group.  Edges colored green span across these groups. 

\begin{figure}
	\includegraphics[width=8cm]{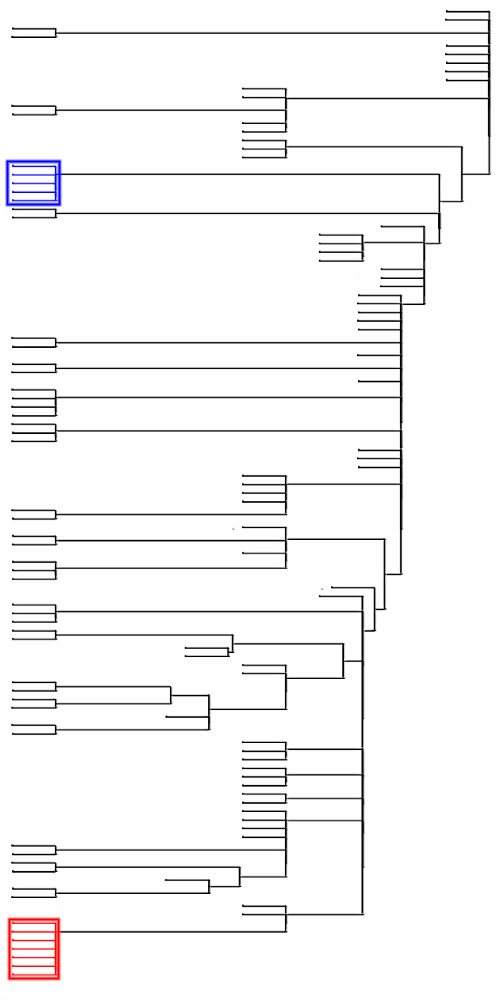}
	\caption{Hierarchical Clustering of Supreme Court Decisions, 1820}
\end{figure}

\begin{figure}
	\includegraphics[width=8cm]{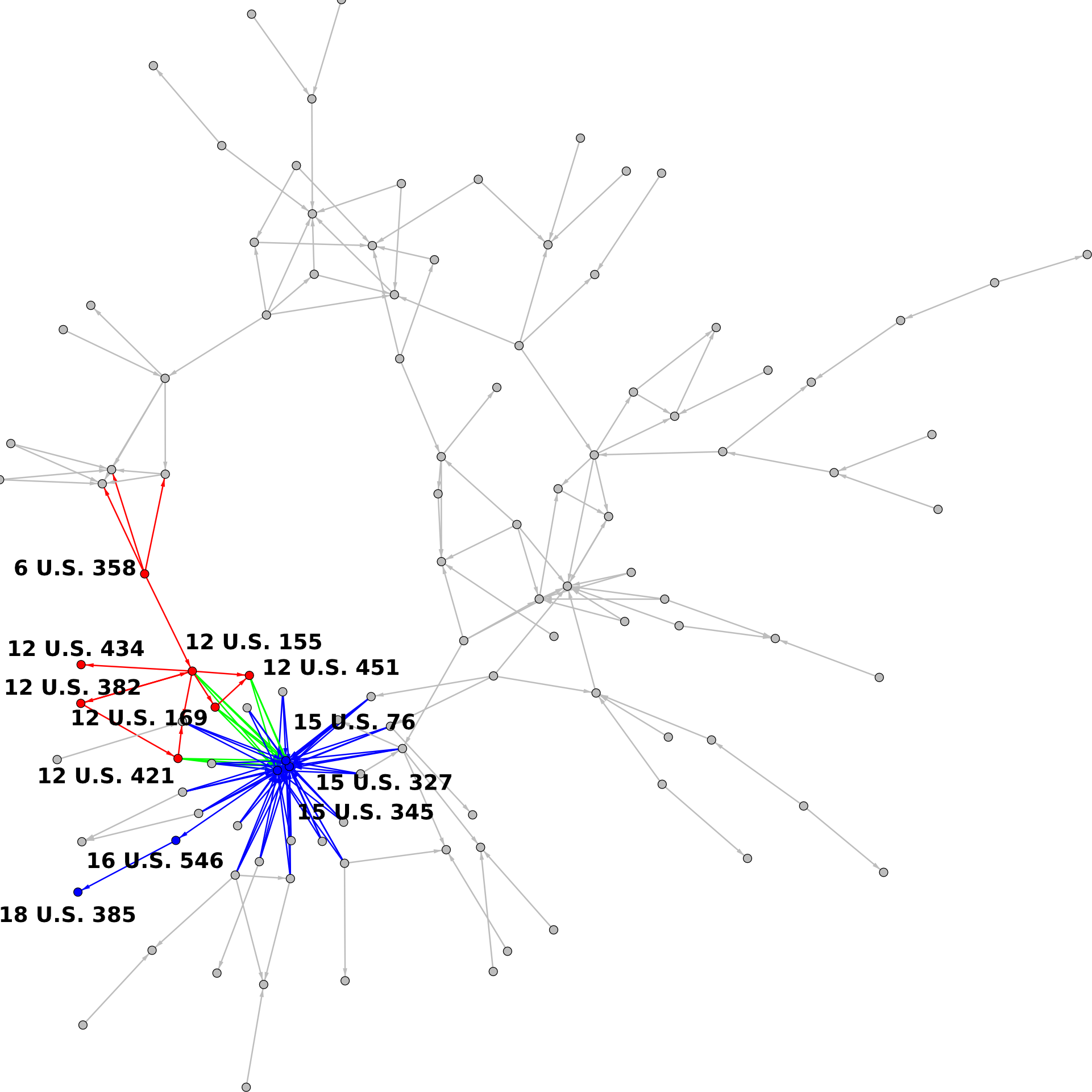}
	\caption{Largest Weakly Connected Component of the Network of Supreme Court Decisions, 1820}

\end{figure}

Both of these colorized clusters engage questions related to admiralty (\footnote{It is important to note that some scholars carve a distinction between admiralty (maritime and private international law) and the laws of the sea (public international law).  For purposes of characterizing the topical domain, however, we believe it is appropriate to broadly identify these as reasonably related to admiralty.}). While not a major focus of the docket of the modern court, the early court elaborated a number of important legal concepts through the lens of these admiralty decisions.  However, despite their general substantive relatedness, these two clusters of cases engage different substantive sub-questions, and are thus appropriately divided into separate clusters.  For example, the red group of cases engages questions of presidential power and the laws of war, as well as general interpretations of the Prize Acts of 1812.  Meanwhile, the blue cluster engages questions surrounding tort liability, jurisdiction, and the burden of proof.  

\section{Conclusion}
We present a novel conception of distance for the class of dynamic citation networks that has trivial implementation and runtime.  We successfully apply our sink approach to not only a theoretical model but also to the citation network of the first quarter-century of United Supreme Court decison. Our method obtains substantively meaningful clustering and is less susceptible to some issues that arise within a high-dimensional citation space. 

Although the substantive application presented here focuses on the decisions of the U.S. Supreme Court, the applicability of this method is likely not limited to judicial citations.  For instance, one could imagine tracing the spread of innovation using sink clustering of the patent citations, or the spread of ideas in an analysis of academic articles.  In future work, we hope to apply this method to such domains, including dimensional data where available.
\\
\section{Acknowledgments}
We would like to thank the University of Michigan Center for the Study of Complex Systems (CSCS) for providing both computational resources and a fruitful research environment.


\begin{thebibliography}{10}
\expandafter\ifx\csname url\endcsname\relax
  \def\url#1{\texttt{#1}}\fi
\expandafter\ifx\csname urlprefix\endcsname\relax\def\urlprefix{URL }\fi

\bibitem{BangJensen2000}
J. Bang-Jensen, G. Gutin. \textit{Digraphs: Theory, Algorithms and Applications}, London: Springer-Verlag, 2000.

\bibitem{Barabasi1999}
A.-L. Barabasi, R. Albert, \textit{Emergence of scaling in random networks}.  Science 286: 509–512, 1999.

\bibitem{Bollobas2003}
B. Bollobas, C. Borgs, J. T. Chayes, O. Riordan. \textit{Directed scale-free graphs}, Proceedings of the Fourteenth Annual ACM-SIAM Symposium on Discrete Algorithms, 2003.

\bibitem{Bommarito2009}
M. J. Bommarito II, D. M. Katz, J. Zelner.  \textit{On the Stability of Community Detection Algorithms on Longitudinal Citation Data}, 2009 (available at http://arxiv.org/abs/0908.0449).

\bibitem{Fowler2007}
J. H. Fowler, T. R. Johnson, J. F. Spriggs II, S. Jeon, P. J. Wahlbeck.  \textit{Network Analysis and the Law: Measuring the Legal Importance of Precedents at the U.S. Supreme Court}, 15 Pol. Analysis, 324, 2007.

\bibitem{Fowler2008}
J. H. Fowler, S. Jeon.  \textit{The Authority of Supreme Court Precedent}, Social Networks 30, 16-30, 2008.

\bibitem{Jaccard1901}
Paul Jaccard.  \textit{Etude comparative de la distribution florale dans une portion des Alpes et des Jura}, 1901.
 
\bibitem{Karrer2009}
B. Karrer, M.~E.~J.~ Newman. \textit{Random acyclic networks}, Phys. Rev. Lett. 102, 128701, 2009.

\bibitem{Leicht2007}
E. A. Leicht, G. Clarkson, K. Shedden, M.E.J. Newman. \textit{Large-scale structure of time evolving citation networks}, Eur. Phys. J. B 59, 75–83, 2007.

\bibitem{Liu2009}
T. Liu, X. Zheng, J. Wang. \textit{Prediction of protein structural class using a complexity-based distance measure}, 	Amino Acids, 2009.

\bibitem{Murtagh1983}
F. Murtagh. \textit{A Survey of Recent Advances in Hierarchical Clustering Algorithms}, The Computer Journal, 1983.

\bibitem{Newman2004}
M.~E.~J.~ Newman, M. Girvan.  \textit{Finding and evaluating community structure in networks}, Physical Review E 69, 2004. 

\bibitem{Newman2008}
M.~E.~J.~ Newman.  \textit{The physics of networks}, Physics Today, 2008.

\bibitem{Porter2009}
M. A. Porter, J.-P. Onnela, and P. J. Mucha. \textit{Communities in Networks}, to appear in Notices of the American Mathematical Society (October 2009 issue).

\bibitem{Post2000}
D. Post, M. Eisen.  \textit{How Long is the Coastline of Law? Thoughts on the Fractal Nature of Legal Systems},  Journal of Legal Studies, Vol. 29, p. 545, 2000.

\bibitem{Smith2007}
T. Smith. \textit{The Web of the Law}, San Diego L. Rev. 44, 309, 2007.

\end{thebibliography}
\end{document}